\documentclass{iaus}
\usepackage{graphicx}

\title[IAU 268] 
{Rotational Mixing and Lithium Depletion}

\author[M. H. Pinsonneault]   
{M. H. Pinsonneault$^1$}

\affiliation{$^1$Ohio State University, Dept. of Astronomy 140 W. 18th
  Ave. Columbus, OH 43210 USA \\ email: {\tt pinsonneault.1@osu.edu} }

\pubyear{2010}
\volume{268}  
\jname{Light elements in the Universe}
\editors{C. Charbonnel, M. Tosi, F. Primas \& C. Chiappini, eds.}
\begin{document}

\maketitle

\begin{abstract}
I review basic observational features in Population I stars which strongly implicate
rotation as a mixing agent; these include dispersion at fixed
temperature in coeval populations and main sequence lithium depletion
for a range of masses at a rate which decays with time.  New
developments related to the possible suppression of mixing at late
ages, close binary mergers and their lithium signature, and an
alternate origin for dispersion in young cool stars tied to radius
anomalies observed in active young stars are discussed.
I highlight uncertainties in models of Population II lithium depletion
and dispersion related to the treatment of angular momentum
loss.  Finally, the origins of rotation are tied to conditions in the
pre-main sequence, and there is thus some evidence that enviroment and
planet formation could impact stellar rotational properties.  This may
be related to recent observational evidence for cluster to cluster
variations in lithium depletion and a connection between the presence
of planets and stellar lithium depletion.
 
\end{abstract}

\firstsection 
\section{Introduction}

Lithium is an extraordinarily sensitive diagnostic of stellar
structure and evolution.  The observed lithium abundances in stars,
not surprisingly, reveal an extremely complex
picture, and it can sometimes be difficult to remember why rotational
mixing is a useful framework for interpreting this data.  I therefore
begin by briefly summarizing the case for rotation as the physical
ingredient responsible for light element depletion in stara.

\subsection{Evidence for Rotational Mixing}

 The first and most important point is that stellar rotation is capable of
driving mild envelope mixing at the observationally required rates 
(\cite[Pinsonneault et al. 1989]{PKDS1989}.)  Rotation induces
a departure from spherical symmetry which generates meridional
circulation currents, and both structural evolution and angular
momentum loss from magnetized winds generate shears which can drive
mild turbulence.  Lithium is easily destroyed in stellar interiors,
and such mild mixing can therefore generate surface lithium depletion.

This leads directly to a second important feature of rotational mixing
which is observationally required: namely, stars which rotate at
different rates will have different mixing histories.  Rapid rotators
experience stronger torques and larger shears than slow rotators, and
they also are less spherical.  It is therefore a basic prediction of
rotational mixing that there should be a dispersion in mixing rates
which can manifest itself as a dispersion in lithium at fixed mass,
composition, and age. Lithium is observed to have a significant
dispersion in many clusters (see \cite[Pinsonneault 1997]{P1997} for a
theoretical review and \cite[Sestito \& Randich 2005]{SR2005} for a more
recent observational synthesis) while other elements in open clusters are very
uniform (\cite[Paulson et al. 2003]{PSC2003}). Other mechanisms, such
as gravity waves and microscopic diffusion, can generate depletion but
not dispersion, so this observed feature allows us to discriminate
between physical processes.

Finally, both the mass dependence and time dependence of the observed
depletion pattern strongly implicate rotationally driven mixing as the
culprit.  Rotation declines with age, and so does lithium depletion.
By contrast, processes such as gravitational settling tend to be more
independent of age, or even increase in rate as stars get older.
Rotational mixing also extends through stellar envelopes, and as a
result it can simultaneously mix different elements and be present in
stars with very different surface convection zone depths.  We observe
lithium depletion in all low mass open cluster stars, which would not
be expected if lithium depletion were a phenomenon confined to the
convection zone boundary.  This does not rule out interesting 
interactions with other physics processes,
such as magnetic or wave-driven angular momentum transport (see the
contribution by Talon in these proceedings), but it
does require rotation as a component of the solution.

However, the physics of stellar angular momentum evolution is
extremely challenging, and it has proven difficult to develop a
rigorous physical model. This has led to a sort of stasis in our
understanding of phenomena such as rotational mixing.  Fortunately,
there have been positiive developments, which I summarize below, which
reveal a dynamic and more complete picture of stellar evolution.  In
Section 2 recent advances in our understanding of angular momentum
evolution are reviewed; Section 3 then discusses three areas where
there are either new observational or theoretical features in stellar
lithium depletion.  A discussion of some recent developments is given
in section 4.

\section{Angular Momentum Evolution}

Stellar rotation is an initial value problem, and the initial
conditions are set by the details of the stsr formation process.  The
angular momentum distribution is subsequently modified by angular
momentum loss (via star-disk interactions) and internal angular
momentum transport.  The physics of the latter is vigorously debated
in the literature, with three distinct mechanisms (hydrodynamic,
wave-driven, and magnetic) all being in principle important.
Rotational mixing is a natural byproduct of angular momentum transport
in stellar radiative interiors, especially from hydrodynamic
mechanisms.  This is a rich field, so I will summarize the main
developments relevant for rotational mixing.

Stars appear at the deuterium-burning birthline (\cite[Stahler
  1988]{S1988}) with a range of rotation rates, typically well below
  that expected for accretion from a Keplerian disk.  The currently
  favored explanation is that magnetic coupling between the protostar
  and the accretion disk regulates the rotation (\cite[Shu et
  al. 1994]{SAL1994}.)  In this framework, the initial rotation rate
  can be thought of as related to the mass accretion rate in the early
  hydrodynamic stages of star formation.  However, the predicted
  rotation rates on the main sequence are both too rapid and too
  uniform if models with the observed rotation rates are evolved to
  the main sequence, even if torques from solar-like winds are
  included.

However, if a coupling between protostars and their accretion disks
exists, the initial spread of rotation rates can be amplified and
stars can reach the main sequence as relatively slow rotators.  Much
observational work has also been invested in the question of star-disk
coupling, with a diversity of results largely centered around the
proper choice of disk proxies and disentangling evolutionary effects.
However, recent Spitzer studies (\cite[Rebull et al. 2006]{RSMHH2006})
have provided strong evidence for a relationship between rotation and
the presence of disks.  This may reflect a coupling between the
protostar and accretion disk similar to that operating at the earlier
stages, or it could be induced by an enhanced stellar wind tied to
accretion.  In either case, the lifetime of accretion disks and their
degree of coupling to the parent star is crucial for establishing the
main sequence rotation.  Rotation is therefore now perceived as a
product of environment, and this raises the interesting possibility
that rotational mixing may also depend on where a star was born or on
how the accretion disk evolved.

There are also now very large databases of stellar rotation periods,
ranging from star forming regions (\cite[Rodriguez-Ledesma et
  al. 2009]{RME2009}) to extensive open cluster surveys such as the
Monitor program(\cite[Irwin et al. 2009]{I2009}) and transit studies
such as the one which yielded a large database of rotation periods
in the 550 Myr system M37(\cite[Hartmann et al. 2009]{H2009}).  The
latter study in particular indicates the ability of modern campaigns
to infer rotation periods caused by spot modulation for large stellar
samples at small amplitude.

These samples can in turn be used to
reconstruct the angular momentum evolution of stellar populations, in
particular the dependence of angular momentum loss on rotation rate
and mass, as well as the coupling timescale between core and envelope
(e.g. \cite[Irwin et al. 2007]{I2007}, \cite[Denissenkov et
  al. 2009]{D2009}.  Different groups agree on the essential
features.  Angular momentum loss scales as the rotation rate cubed at
low rotation rates, then saturates at a threshold which decreases as
mass descreases.  The net effect is that lower mass stars take longer
to spin down and longer for their rotation rates to converge.  The
cores of the slowly rotating population couple to their envelopes with
a timescale of order 100 Myr, while rapidly rotating stars appear to
be more strongly coupled.  These results are consistent with
helioseismic data indicating that the rotationa of the solar interior is strongly
coupled to that of the surface convection zone.

This combination of theoretical advances and improved rotational data
and empirical constraints therefore has significant promise for more
robust rotational mixing predictions in the future.

\section{Lithium Depletion and Rotational Mixing Revisited}

The basic rotational mixing picture can be simply defined.  Stars
experience a mass-dependent pre-main sequence lithium burning epoch,
which ends when they develop substantial radiative cores.  They then
experience rotational mixing on the main sequence, induced either by
shears generated by angular momentum loss or their departure from
spherical symmetry.  As the stars spin down the rate of lithium
depletion decreases.  This overall picture is reasonable, but a number
of phenomena defy easy categorization within it.  This is in large
part because of the interaction of rotation with other phenomena
typically neglected in stellar models.  Below are three examples.

\subsection{Structural Effects of Starspots}

There is a striking dispersion in lithium abundances among late-type
stars in young open clusters; the Pleiades is the clearest example
(\cite[Soderblom et al. 1993]{S1993}.)  This trend is not expected
from rotational mixing in such young stars, and the relative effect is
also the opposite of the one expected: namely, the least depleted
stars are the most active and heavily spotted.  Much subsequent work
has focused on whether the dispersion is real or induced by the
heavily spotted nature of the stars in question; the model atmospheres
used to interpret the data typically neglect the large changes in the
strength of the lithium feature which would be associated with a
substantial fraction of the surface covered with cool spots with ample
neutral lithium.  However, in recent work (\cite[King et
  al. 2009]{KSHP2009}) we found that the scatter in K I was much less
than the scatter in Li I, indicating that the bulk of the dispersion
is real.  The likely origin in our view is actually a different
mechanism altogether, and it is motivated by recent data on radius
anomalies in active stars from interferometric and eclipsing binary studies.

\begin{figure}[b]
\begin{center}
 \includegraphics[width=3.4in]{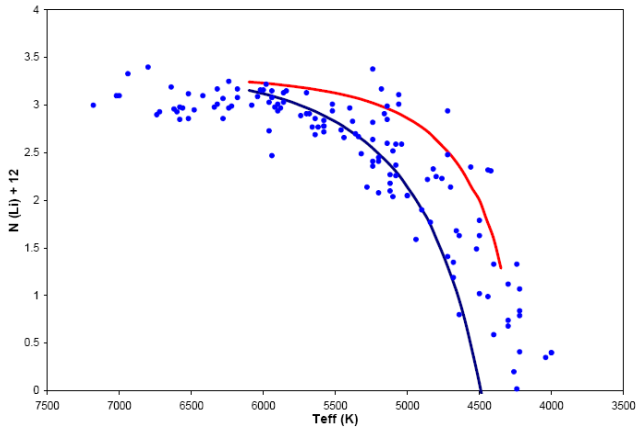} 
 \caption{Lithium abundances as a function of effective temperature in
 the Pleiades cluster compared with standard models (lower line) and
 models which were inflated during the pre-main sequence lithium
 depletion epoch (upper line).}
   \label{fig1}
\end{center}
\end{figure}

Eclipsing binaries such as YY Gem (\cite[Torres \& Ribas 2002]{TR2002})
were found to have radii significantly larger than those predicted by interiors theory.
Subsequent work traced out a radius anomaly pattern.  More recent
interferometric data permits the measurement of radii for inactive
field stars, which are found to be in accord with theoretical
predictions (\cite[Demory et al. 2009]{Dem2009}.)  High activity, such
as that found in tidally synchronized short-period binaries, therefore
appears to puff up stars.  A similar effect during the fully
convective protostellar phase would reduce the degree of pre-main
sequence lithium depletion; if this varied from star to star it could
generate a dispersion with characteristics remarkably like the data.
This is illustrated in Fig.\,\ref{fig1}., where Pleiades data from Soderblom et
al. 1993 is compared with standard stellar models (lower line) and
models with a radius inflated by 10 percent, the level inferred in
highly active stars (upper line).  In addition to being an attractive solution for
a longstanding problem, this leads to an interesting insight.  Stellar
activity is not a mere detail; it can impact the entire structure
of a star and change its mixing history.

\subsection{Blue Stragglers and Halo Lithium Depletion}

The lithium depletion pattern in metal-poor stars poses a different
problem; the majority of stars exhibit little dispersion and the
observed abundances appear to be nearly independent of surface
temperature or metallicity.  One striking counter-example is the
presence of a small but real population of highly depleted stars
(\cite[Thorburn 1994]{T1994}.)  In recent work on blue stragglers we
find that a population of sub-turnoff merger products, presumably
highly lithiumn depleted, is predicted to arise from such mergers; the
number expected is close to that observed in halo stars
(\cite[Andronov, Pinsonneault \& Terndrup 2006]{APT2006}.)  This
confirms the suggestion in \cite[Ryan et al. (2001)]{RBKR2001} that
these highly depleted stars should not be regarded as the tail of a
rotational mixing distribution, but rather that they have a distinct
origin.

This does not, however, require that mixing be absent in halo stars.
The nature of the dispersion predicted depends on both the
distribution of initial conditions and the angular momentum loss
history.  At present we can only extrapolate Population I conditions
to Population II stars.  Their torques and initial conditions could
very well have been different; for example, planet formation may be
less common in them, and this could impact the distribution of
accretion disk coupling timescales (see Section 4.)  Future work on
the activity properties of tidally synchronized halo stars may prove
diagnostics of the braking law, while we may need rotation data in
more metal poor outer disk systems to test the metallicity dependence
of the initial conditions.

\subsection{Interaction of Diffusion and Rotational Mixing}

Microscopic diffusion (or gravitational settling and thermal
diffusion) is a basic physical process expected to occur in stars,
typically over a very long timescale.  The net effect is that heavy
species tend to sink relative to light ones, although radiation
pressure can drive some heavier elements upwards in sufficiently thin
surface convection zones (\cite[Michaud 1970]{M1970}.  There is clear
evidence for diffusion in the Sun (\cite[Bahcall \& Pinsonneault
  1992]{BP1992}), both in the sound speed profile and in the detection
of a surface helium abundance lower than that initially required to
reproduce the solar luminosity.  

Diffusion can induce lithium depletion directly, but it also has interesting
interactions with rotational mixing (see \cite[Richard et
  al. 1996]{Rich1996} for a nice example in the solar context.)  
Gravitational settling operates over shorter timescales for thinner convection zones, and it produces
mean molecular weight gradients at the base of the surface convection
zone.  It is energetically unfavorfable to mix in the presence of a
mu gradient, and mixing can erase composition gradients; there is thus
a natural competition between the two processes.  Furthermore, the
timescale for mixing increases with age, while the 
timescale for settling changes very slowly (and tends to decrease as
stars evolve to higher effective temperatures.)  One might therefore
expect rotational mixing to predominate earlier while diffusion
suppresses mixing at later ages, and for this interaction to depend on
mass and composition.  This may be related to the apparent
stalling of lithium depletion in older open clusters (discussed by
Randich in these proceedings), and could be an additional source of
lithium depletion in halo stars as well.  Recent evidence for settling
in multiple elements (Korn, these proceedings) of globular cluster
stars provides evidence that diffusion sets in for older stars; an
earlier epoch of depletion is certainly permitted by theory, although
establishing this observationally will require additional work as
discussed in the angular momentum evolution section.  Observations of
multiple elements, as already done in globulars, could be used to
establish the diffusion signature in old open clusters and the
interaction between mixing and separation.

\section{Future Directions}

In closing I'd like to note some other wrinkles which may prove
important for understanding lithium depletion: differences in depletion
patterns from cluster to cluster (see the presentation by Randich) 
and an apparent excess lithium depletion in stars which host planets.  
data (see the talks by Israelian and, for a contrary view, Melendez.)
Both can be interpreted in the framework where stellar rotation
properties are determined by interactions between protostars and
accretion disks.  In dense stellar environments the timescale for
interactions can be comparable to the lifetimes inferred for accretion
disks, raising the possibility that stars born in such regions might
have a different distribution of disk lifetimes than stars born in
loose associations.  This hypothesis is testable in the measured
rotation rates of young systems, and this is an important potential
effect (especially if we use clusters as an evolutionary sequence!)
which needs to be explored.

The recent report that stars with planets have excess lithium
depletion (\cite[Israelian et al. 2009]{Isr2009}) may be a fascinating
example of how the formation of planets can impact the properties of
stars.  \cite[Bouvier (2008)]{Bou2008} has proposed a linkage, arguing
that systems with planets should have long-lived accretin disks.
These in turn become slow rotators, with large relative shears, which
in turn could drive excess mixing.  He thus argued that there may be a
connection between lithium overdepletion and planet formation.  Such a 
link is certainly plausible, but the opposite correlation
appears to be required by rotational mixing.  More rapid rotators
experience larger absolute torques (and in any case subsequently
evolve to become slow rotators, thus in effect adding the mixing from
the rapid to that of the slow phase.)  They also experience larger
departures from spherical symmetry; both imply stronger mixing.  
However, the rotation is set primarily by the coupling between star
and disk, not necessarily in the disk lifetime itself, and this may
explain the apparent contradiction between ``massive disk required for
planets'' and ``weak star-disk interaction required for rapid rotation
and lithium depletion.''   This avenue may prove promising to explore,
and it would be a delightful turn of events if the planetary tail
could wag the stellar dog.

\end{document}